\documentclass[aps,pra,twocolumn,showpacs,groupedaddress,superscriptaddress,nofootinbib]{revtex4}
\usepackage{bbm,amsmath,amssymb,amsbsy,graphicx,subfigure,txfonts}

\usepackage{hyperref,color}
\usepackage[latin1]{inputenc}
\newcommand*{\bfrac}[2]{\genfrac{.}{.}{0pt}{}{#1}{#2}}

\newcommand{\be}{\begin{equation}}
\newcommand{\ee}{\end{equation}}
\newcommand{\ben}{\begin{eqnarray}}
\newcommand{\een}{\end{eqnarray}}
\newcommand{\bes}{\begin{subequations}}
\newcommand{\ees}{\end{subequations}}

\newcommand{\tr}{\text{tr}}
\newcommand{\gr}[1]{\boldsymbol{#1}}
\newcommand{\ket}[1]{|#1\rangle}
\newcommand{\bra}[1]{\langle#1|}

\newcommand{\eq}[1]{Eq.~(\ref{#1})}

\newcommand{\sig}{{\gr\sigma}}
\newcommand{\gam}{\boldsymbol{\gamma}}

\newtheorem{theorem}{Theorem}

\begin{document}
\title{Measuring Gaussian Quantum Information and Correlations Using the R\'{e}nyi Entropy of Order $\boldsymbol 2$}

\date{August 17, 2012}
\author{Gerardo Adesso}
\affiliation{School of Mathematical Sciences, University of Nottingham,
University Park,  Nottingham NG7 2RD, U.K.}

\author{Davide Girolami}
\affiliation{School of Mathematical Sciences, University of Nottingham,
University Park,  Nottingham NG7 2RD, U.K.}

\author{Alessio Serafini}
\affiliation{Department of Physics \& Astronomy, University College London, Gower Street, London WC1E 6BT, U.K.}


\begin{abstract}
We demonstrate that the R\'{e}nyi-$2$ entropy provides a natural measure of information for any multimode Gaussian state of quantum harmonic systems, operationally linked to the phase-space Shannon sampling entropy of the Wigner distribution of the state. We prove that, in the Gaussian scenario, such an entropy satisfies the strong subadditivity inequality, a key requirement for quantum information theory. This allows us to define and analyze measures of Gaussian entanglement and more general quantum correlations based on such an entropy, which are shown to satisfy relevant properties such as monogamy.
\end{abstract}


\maketitle

In quantum information theory, the degree of information contained in a quantum state $\rho$ is conventionally quantified via the von Neumann entropy of the state ${\cal S}(\rho) = - \text{tr}(\rho \ln \rho)$, that is {\it the} direct counterpart to the Shannon entropy in classical information theory \cite{zykbook}. The most fundamental mathematical implications and physical insights in quantum information theory, ranging from the Holevo bound \cite{holevobound} all the way to the whole genealogy of quantum communication protocols \cite{merging,mother}, rely on a key property satisfied by the von Neumann entropy, the strong subadditivity inequality \cite{proofsss,wehrl}
\begin{equation}\label{eq:ssub}
{\cal S}(\rho_{AB}) + {\cal S}(\rho_{BC}) \geq {\cal S}(\rho_{ABC}) + {\cal S}(\rho_{B})\,,
\end{equation}
for an arbitrary tripartite state $\rho_{ABC}$.
The strong subadditivity inequality implies in particular that the mutual information
\begin{equation}\label{eq:mutual}
{\cal I}(\rho_{A:B}) = {\cal S}(\rho_A)+{\cal S}(\rho_B)-{\cal S}(\rho_{AB})\,,
\end{equation}
which measures the total correlations between the subsystems $A$ and $B$ in the bipartite state $\rho_{AB}$,
is always nonnegative.  However, in classical as well as quantum information theory, several other entropic quantities have been introduced and studied. In particular, R\'{e}nyi-$\alpha$ entropies \cite{renyi} are an interesting family of additive entropies, whose interpretation is related to derivatives of the free energy with respect to temperature \cite{baez}, and which have found applications especially in the study of channel capacities \cite{renyicapacity}, work value of information \cite{workvalue}, and entanglement spectra in many-body systems \cite{renyispectrum}. They are defined as
\begin{equation}\label{eq:ren}
{\cal S}_\alpha(\rho) = (1-\alpha)^{-1} \ln \text{tr} (\rho^\alpha)\,,
\end{equation}
and reduce to the von Neumann entropy in the limit $\alpha \rightarrow 1$.
R\'{e}nyi entropies are powerful quantities for studying quantum correlations in multipartite states. For a bipartite pure state $\rho_{AB}$, any of the entropies in Eq.~\eqref{eq:ren} evaluated on the reduced density matrix of one subsystem only, say $\rho_A$, is an entanglement monotone \cite{vidal}, dubbed R\'{e}nyi-$\alpha$ entanglement \cite{barrylagallina}. Any such measure can be extended to mixed states via conventional convex roof techniques \cite{entanglement}. Using the $\alpha=2$ instance, ${\cal S}_2(\rho) = - \ln \text{tr} (\rho^2)$, the ensuing R\'{e}nyi-$2$ measure of entanglement has been defined and proven to satisfy the important `monogamy' inequality \cite{ckw,osborne} for multiqubit states \cite{marcio,barrylagallina}.
 Nonetheless, R\'{e}nyi-$\alpha$ entropies for $\alpha \neq 1$ are not in general subadditive \cite{zykbook}; this entails, e.g., that if one replaces ${\cal S}$ by ${\cal S}_{\alpha \neq 1}$ in (\ref{eq:mutual}), the corresponding quantity can become negative, i.e., meaningless as a correlation measure; see \cite{lptp} for explicit two-qubit instances of this fact when  $\alpha=2$.

In this Letter, we focus our attention on multimode quantum harmonic oscillators, and analyze the informational properties of general Gaussian states as measured by the R\'{e}nyi-$2$ entropy. Gaussian states constitute versatile resources for quantum communication protocols with continuous variables \cite{brareview,book} and are important testbeds for investigating the structure of quantum correlations \cite{ourreview}, whose role is crucial in fields as diverse as quantum field theory, solid state physics, quantum stochastic processes, and open system dynamics. Gaussian states naturally occur as ground or thermal equilibrium states of any physical quantum system in the `small-oscillations' limit \cite{ciracreview,extra}, and can be very efficiently engineered, controlled and detected in various experimental setups, including light, atomic ensembles, trapped ions, nano-/opto-mechanical resonators, and hybrid interfaces thereof \cite{book}.
Given the special role played by Gaussian states and operations, which are formally `extremal' with respect to several quantum primitives \cite{extra}, a Gaussian-only theory of quantum information is actively pursued \cite{jenswolf,GQI}. In particular, a key conjecture which ---despite its disproof for general channels \cite{hastings}--- still stands within the Gaussian scenario, is the so-called minimum output entropy conjecture for bosonic channels \cite{jenswolf,GQI,giovannetti,serafozzinazi}. Interestingly, the conjecture is verified (for all phase-insensitive channels) if using R\'{e}nyi-$\alpha$ entropies with $\alpha \ge 2$ \cite{giovannetti2}, but to date is resisting an analytical proof for  $\alpha \rightarrow 1$ \cite{lloydcacconecerfnew}.  This somehow raises the question whether the von Neumann entropy is indeed the most natural one within the Gaussian scenario \cite{extremal}.

Here we show that the R\'{e}nyi-$2$ entropy should be regarded as a specially meaningful choice to develop a Gaussian theory of quantum information and correlations.
We prove that the R\'{e}nyi-$2$ entropy ${\cal S}_2$ satisfies the strong subadditivity inequality (\ref{eq:ssub}) for all Gaussian states, is operationally linked to the Shannon entropy of Gaussian Wigner distributions (providing a natural measure of state distinguishability in phase space), and can be employed to define valid measures of  Gaussian entanglement \cite{geof,ordering} and general  {\it discord}-like  quantum correlations \cite{zurek,henvedral,modireview,adessodatta,giordazzo}, whose monogamy properties \cite{ckw,osborne,contangle,hiroshima,georgie,arequantum} we investigate in detail. In particular, we obtain a truly {\it bona fide} measure of genuine tripartite entanglement for three-mode Gaussian states, based on ${\cal S}_2$.
Our study allows us to explore within a unified framework the various facets of nonclassicality in the Gaussian realm.

\noindent{\bf R\'{e}nyi-$\bf 2$ entropy for Gaussian states.}---
We consider a $n$-mode continuous variable system; we collect the quadrature operators in the vector $\boldsymbol{\hat R} = (\hat q_1, \hat p_1, \hat q_2, \hat p_2, \ldots, \hat q_n, \hat p_n)^{\sf T} \in \mathbb{R}^{2n}$ and write the canonical commutation relations compactly as $[\hat R_j, \hat R_k] = i \left(\boldsymbol\omega^{\oplus n}\right)_{j,k}$ with $\boldsymbol\omega = {{\ 0 \ \ 1} \choose {-1 \ 0}}$ being the symplectic matrix. A Gaussian state $\rho$ is described by a positive, Gaussian-shaped  Wigner distribution in phase space,
\begin{equation}
\label{eq:wigner}
W_\rho(\boldsymbol{\xi}) = \frac{1}{\pi^n \sqrt{\det{\gam}}} \exp\big(-\boldsymbol{\xi}^{\sf T} \boldsymbol{\gamma}^{-1} \boldsymbol{\xi}\big)\,,
\end{equation}
where $\boldsymbol{\xi} \in \mathbb{R}^{2n}$, and $\boldsymbol{\gamma}$ is the covariance matrix (CM) of elements $\gamma_{j,k}=\text{tr}[\rho \{\hat{R}_j,\hat{R}_k\}_+]$, which (up to local displacements) completely characterizes the Gaussian state $\rho$.

Let us now evaluate different entropic quantities on the Gaussian state $\rho$. All the measures defined below are invariant under local unitaries, so we will assume our states to have zero first moments, $\langle \boldsymbol{\hat R}\rangle = \boldsymbol{0}$, without loss of generality. The purity is easily computed as $\tr\,\rho^2 = (2\pi)^n \int_{\mathbb{R}^{2n}}  W^2_\rho(\boldsymbol{\xi}) \ {\rm d}^{2n} \boldsymbol{\xi} = (\det\boldsymbol{\gamma})^{-\frac12}$.
Hence the R\'{e}nyi-$2$ entropy of an arbitrary $n$-mode Gaussian state is
\begin{equation}\label{eq:renyg}
{\cal S}_2(\rho) =  \frac12 \ln (\det \boldsymbol{\gamma})\,,
\end{equation}
ranging from $0$ on pure states ($\det \boldsymbol{\gamma}=1$) and growing unboundedly with increasing mixedness (i.e., temperature) of the state.
The von Neumann entropy is instead a more complicated function that depends on the local temperatures of all the $n$ normal modes of $\rho$, that is, on the full symplectic spectrum of $\boldsymbol\gamma$ \cite{holevowerner,serafozzididattico,ourreview,extremal}. On the other hand, since the Wigner function is a valid probability distribution providing a fully equivalent description of $\rho$, we can alternatively compute the Shannon entropy of $W_\rho$ to obtain an alternative quantifier of the informational content of any Gaussian state. Such an entropy has a clear interpretation in terms of phase-space sampling via homodyne detections \cite{vecchiocavaliere}. The continuous (Boltzmann--Gibbs--)Shannon entropy for a probability distribution $P(x)$ is defined as $H(P) = - \int P(x) \ln P(x)\; {\rm d}x $ \cite{shannon}. It can be shown that \cite{stratonovich} (see the Appendix \cite{epaps} for a derivation)
\begin{eqnarray}\label{eq:sampling}
H(W_\rho) &=& -\int_{\mathbb{R}^{2n}}  W_\rho(\boldsymbol{\xi}) \ln [W_\rho(\boldsymbol{\xi})]\ {\rm d}^{2n} \boldsymbol{\xi} \nonumber \\
&=& {\cal S}_2(\rho) + n(1+\ln \pi)\,.
\end{eqnarray}
Interestingly, we see that this sampling entropy coincides (modulo the additional constant) with the R\'{e}nyi-$2$ entropy rather than  the von Neumann one for Gaussian states.
We can now take an extra step and introduce a `relative sampling entropy', to quantify the phase-space distinguishability of two $n$-mode Gaussian quantum states $\rho_1$ and $\rho_2$ (with CMs $\boldsymbol{\gamma}_1$ and $\boldsymbol{\gamma}_2$, respectively), defined as the relative Shannon entropy (also known as Kullback-Leibler divergence \cite{wehrl,zykbook}) between their respective Wigner distributions $W_{\rho_1}$ and $W_{\rho_2}$, yielding \cite{epaps}
\begin{eqnarray}\label{eq:relent}
H\big(W_{\rho_1}\|W_{\rho_2}\big) &\doteq& \int_{\mathbb{R}^{2n}}  W_{\rho_1}(\boldsymbol{\xi}) \ln\left(\frac{W_{\rho_1}(\boldsymbol{\xi})}{W_{\rho_2}(\boldsymbol{\xi})}\right)\ {\rm d}^{2n} \boldsymbol{\xi}  \\
&=& \frac12 \left[\ln\left(\frac{\det\boldsymbol{\gamma}_2}{\det\boldsymbol{\gamma}_1}\right) + \tr\left(\gam_1 \gam_2^{-1}\right)\right]-n\,.  \nonumber 
\end{eqnarray}
Let us evaluate Eq.~(\ref{eq:relent}) when $\rho_1 \equiv \rho_{AB}$ is a generic Gaussian state of a composite system, partitioned into two subsystems $A$ and $B$ (of $n_A$ and $n_B$ modes respectively, with $n_A+n_B=n$), and $\rho_2 \equiv \rho_A \otimes \rho_B$ is  the tensor product of the two marginals of $\rho_{AB}$. Writing the CMs in block form,
\begin{equation}\label{eq:cms}
\gam_1 \equiv \gam_{AB} = \left(\begin{array}{c|c}
\gam_A & {\gr\varsigma}_{AB} \\ \hline
{\gr\varsigma}_{AB}^{\sf T} & \gam_B
\end{array}
\right)\,,\quad
 \gam_2 \equiv \gam_A \oplus \gam_B\,,
\end{equation}
we have $\tr\left(\gam_1 \gam_2^{-1}\right) = 2(n_A+n_B)=2n$, which entails
\begin{eqnarray}\label{eq:remutual}
H\big(W_{\rho_{AB}}\|W_{\rho_A \otimes \rho_B}\big) &=& H(W_{\rho_A}) + H(W_{\rho_B}) - H(W_{\rho_{AB}}) \nonumber \\
&=&\frac12 \ln\left(\frac{\det\boldsymbol{\gamma}_A \det\boldsymbol{\gamma}_B}{\det\boldsymbol{\gamma}_{AB}}\right) \nonumber \\
&=& {\cal S}_2(\rho_A) + {\cal S}_2(\rho_B) - {\cal S}_2(\rho_{AB}) \nonumber \\
&\doteq& {\cal I}_2(\rho_{A:B})\,.
\end{eqnarray}
The above equation defines the `Gaussian R\'{e}nyi-$2$ (GR$2$) mutual information' ${\cal I}_2$ for an arbitrary bipartite Gaussian state $\rho_{AB}$. It follows  that  ${\cal I}_2(\rho_{A:B}) \geq 0$, as it coincides exactly with the Shannon continuous mutual information of the Wigner function of $\rho_{AB}$, which is positive semidefinite. The expression in \eq{eq:remutual}, which is analogous to the more familiar von Neumann one [\eq{eq:mutual}], has thus a precise operational interpretation as the amount of extra discrete information (measured in natural bits) that needs to be transmitted over a continuous variable channel to reconstruct the complete joint Wigner function of $\rho_{AB}$ rather than just the two marginal Wigner functions of the subsystems \cite{shannon,vecchiocavaliere}; in short, ${\cal I}_2(\rho_{A:B})$ measures the total quadrature correlations of  $\rho_{AB}$.

We can then enquire whether a more general property such as the strong subadditivity inequality (\ref{eq:ssub}) ---which would independently imply the nonnegativity of ${\cal I}_2$--- holds  in general for ${\cal S}_2$ in the Gaussian scenario.
Let $\rho_{ABC}$ be a tripartite Gaussian state whose subsystems encompass arbitrary number of modes. Writing its CM in block form as in Eq.~(\ref{eq:cms}), and using the definition (\ref{eq:renyg}), we have the following
\begin{theorem}\label{teo}
The R\'{e}nyi-$2$ entropy ${\cal S}_2$ satisfies the strong subadditivity inequality for all Gaussian states $\rho_{ABC}$,
\begin{equation}\label{eq:ss2}\begin{split}
 {\cal S}_2(\rho_{AB}) + {\cal S}_2(\rho_{BC}) - {\cal S}_2(\rho_{ABC}) - {\cal S}_2(\rho_{B}) \\
\quad= \frac12 \ln \left( \frac{\det\gam_{AB} \det\gam_{BC}}{\det\gam_{ABC}\det\gam_B} \right) \geq 0\,.
\end{split}\end{equation}
\end{theorem}
\noindent {\it Proof.} The result follows by applying a particular norm compression inequality to the CM $\gam_{ABC}$. Given a positive  Hermitian matrix $\gr A \in \mathbb{M}_m$, and given any two index sets $\alpha, \beta \subseteq N=\{1,\ldots,m\}$, the Hadamard-Fisher inequality \cite{hornjohnson} states that
$\det \gr A_{\alpha \cup \beta} \det \gr A_{\alpha \cap \beta} \leq \det \gr A_\alpha \det \gr A_\beta$.
Recalling that any CM $\gam_{ABC}$ is a positive  real symmetric matrix \cite{williamson}, the claim follows upon identifying $\alpha$ with the indices of modes $AB$ and $\beta$ with the indices of modes $BC$. \hfill $\blacksquare$

Beyond its apparent simplicity, Theorem \ref{teo} has profound consequences. It yields that the  core of quantum information theory can be consistently reformulated, within the Gaussian scenario \cite{jenswolf,GQI}, using the simpler and physically natural  R\'{e}nyi-$2$ entropy in alternative to the von Neumann one.
 In the rest of this Letter we will focus on defining GR$2$ quantifiers of entanglement and other correlations for Gaussian states.

\noindent{\bf Gaussian R\'{e}nyi-$\bf 2$ measures of correlations.}---
The GR$2$ entanglement ${\cal E}_2$ can be defined as follows. Given a $n$-mode bipartite Gaussian state $\rho_{AB}$ with CM $\gam_{AB}$,
\begin{equation}\label{eq:GR2_ent}
{\cal E}_2 (\rho_{A:B}) \doteq \inf_{\bfrac{\sig_{AB}:} {\left\{\bfrac{
  \pm i \boldsymbol\omega^{\oplus n}
 \leq\sig_{AB} \le \gam_{AB},}{\det{\sig_{AB}}=1}\right\}}} \frac12 \ln \left(\det \sig_A\right)\,.
\end{equation}
For a pure Gaussian state $\rho_{AB} = \ket{\psi_{AB}}\bra{\psi_{AB}}$, the minimum is saturated by $\sig_{AB}=\gam_{AB}$, so that ${\cal E}_2(\psi_{A:B})= {\cal S}_2(\rho_A) = \frac12 \ln (\det{\gam_A})$, where $\gam_A$ is the reduced CM of subsystem $A$. For a generally mixed state, Eq.~(\ref{eq:GR2_ent}) ---where the minimization is over pure $n$-mode Gaussian states with  CM $\sig_{AB}$ smaller than $\gam_{AB}$---  amounts to taking the Gaussian convex roof of the pure-state R\'{e}nyi-$2$ entropy of entanglement, according to the formalism of \cite{geof,ourreview}. Closed formulae for ${\cal E}_2$ can be obtained for special classes of two-mode Gaussian states \cite{eofg,ordering,3modi} exploiting the same procedure adopted for the Gaussian entanglement of formation \cite{geof,epaps}. Like the latter measure (and all Gaussian convex-roof entanglement measures), it follows from the results of \cite{geof} that the GR$2$ entanglement is in general monotonically nonincreasing under Gaussian local operations and classical communication, and is additive for two-mode symmetric Gaussian states.

We can also introduce a `GR$2$ measure of one-way classical correlations' in the spirit of Henderson and Vedral \cite{henvedral}. We define ${\cal J}_2(\rho_{A|B})$ as the maximum decrease in the R\'{e}nyi-$2$ entropy of subsystem $A$, given a Gaussian measurement has been performed on subsystem $B$, where the maximization is over all Gaussian measurements---i.e.~those  that  map Gaussian states into Gaussian states \cite{adessodatta,giordazzo}.
Any such  measurement on, say,  the $n_B$-mode subsystem $B=(B_1\ldots B_{n_B})$, is described by a positive operator valued measure (POVM)  \cite{fiurasek07} of the form $\Pi_B(\gr{\eta}) = \pi^{-n_B} [\prod_{j=1}^{n_B}  \hat{W}_{B_j}(\eta_j)] \Lambda^\Pi_B [\prod_{j=1}^{n_B}\hat{W}^\dagger_{B_j}(\eta_j)]$ where $\hat{W}_B(\eta_j) = \exp(\eta_j \hat{b}_j^\dagger - \eta_j^\ast \hat{b}_j)$ is the Weyl operator, $\hat{b}_j = (\hat{q}_{B_j} + i \hat{p}_{B_j})/\sqrt2$, $\pi^{-n_B}\int \Pi_B(\gr{\eta}) {\rm d}^{2n_B}\gr{\eta}  = \openone$, and $\Lambda^\Pi_B$ is the density matrix of a (generally mixed) $n_B$-mode Gaussian state with CM $\gr{\Gamma}_B^\Pi$ which denotes the seed of the measurement. The conditional state $\rho_{A|\gr{\eta}}$ of subsystem $A$ after the measurement $\Pi_B(\gr{\eta})$ has been performed on $B$ has a CM $\tilde{\gr\gamma}^{\Pi}_A$ independent of the outcome $\gr{\eta}$ and given by the Schur complement \cite{giedkefiurasekdistill}
\begin{equation}\label{eq:schur}
\tilde{\gr\gamma}^{\Pi}_A = \gam_A-\gr\varsigma_{AB} (\gr\gamma_B+ \gr\Gamma_B^\Pi)^{-1} \gr\varsigma_{AB}^{\sf T}\,,
 \end{equation}
 where the original bipartite CM $\gam_{AB}$ of the $n$-mode  state $\rho_{AB}$ has been written in block form as in (\ref{eq:cms}).
We have then
\begin{equation}\label{eq:J2}
{\cal J}_2(\rho_{A|B}) \doteq \sup_{\gr\Gamma_B^\Pi} \frac12 \ln \left(\frac{\det \gam_A}{\det\tilde{\gr\gamma}^{\Pi}_A}\right)\,.
\end{equation}
The one-way classical correlations ${\cal J}_2(\rho_{B|A})$, with measurements on $A$, can be defined accordingly by swapping $A \leftrightarrow B$.

We can now define a Gaussian measure of quantumness of correlations based on R\'{e}nyi-$2$ entropy. Following the seminal study by Ollivier and Zurek \cite{zurek}, and the recent analysis of Gaussian quantum discord using  von Neumann entropy \cite{giordazzo,adessodatta}, we define the `GR$2$ discord' as the difference between mutual information (\ref{eq:remutual}) and classical correlations (\ref{eq:J2}),
\begin{equation}\label{eq:D2}
\begin{split}
{\cal D}_2(\rho_{A|B}) &\doteq {\cal I}_2(\rho_{A:B})- {\cal J}_2(\rho_{A|B}) \\
&=\inf_{\gr\Gamma_B^\Pi} \frac12 \ln \left(\frac{\det \gam_B \det \tilde{\gr\gamma}^{\Pi}_A}{\det \gam_{AB}}\right)\,.
\end{split}
\end{equation}
For the case of $A$ and $B$ being single modes, that is, $\rho_{AB}$ being a general two-mode Gaussian state, closed formulae can be obtained for Eqs.~(\ref{eq:J2},\ref{eq:D2}) thanks to the results of \cite{adessodatta}. We report them in the Appendix \cite{epaps} for completeness. We remark that Theorem~\ref{teo} is crucial to guarantee the nonnegativity and the faithfulness of the GR$2$ discord \cite{dattanullity,hjpw04,footnotedatta}.
Notice also that $\frac12 {\cal I}_2(\rho_{A:B})={\cal J}_2(\rho_{A|B})={\cal D}_2(\rho_{A|B})={\cal S}_2(\rho_A)$ for pure bipartite Gaussian states $\rho_{AB}$.

A trade-off relation between the entanglement ${\cal E}_2$ and the classical correlations ${\cal J}_2$ can be written for arbitrary tripartite pure Gaussian states $\rho_{ABC}$ \cite{adessodatta}, following Koashi and Winter \cite{koashi}. One can essentially exploit the fact that all possible Gaussian POVMs on $B$ induce all possible Gaussian pure-state decompositions of the subsystem $AC$  (see also \cite{osborne}), which implies
\begin{equation}\label{eq:kw}
{\cal S}_2(\rho_A) = {\cal J}_2(\rho_{A|B}) + {\cal E}_2(\rho_{A:C})\,.
\end{equation}
This relation can be manipulated to express the ``conservation'' of different types of correlations in
a generic pure tripartite Gaussian state $\rho_{ABC}$, along the lines of \cite{fanchini}.

We now look at monogamy properties of the GR$2$ measures. 
For an entanglement monotone $E$ and a $n$-partite state $\rho_{A_1 A_2 \ldots A_n}$, the monogamy relation (choosing party $A_1$ as the focus), which constrains the distribution of bipartite entanglement among different splits, can be written as \cite{ckw} $E(\rho_{A_1:A_2 \ldots A_n}) - \sum_{j=2}^{n} E(\rho_{A_1:A_j}) \geq 0\,.$
The R\'{e}nyi-$2$ entanglement measure \cite{marcio,barrylagallina}, as well as the tangle (squared concurrence) \cite{ckw,osborne}, satisfy this inequality for general $n$-qubit states. A Gaussian version of the tangle (based on squared negativity) has been defined that obeys the inequality for all $n$-mode Gaussian states \cite{hiroshima}. We now show that ${\cal E}_2$ does too.
\begin{theorem}\label{teomono}
The GR$2$ entanglement defined in Eq.~(\ref{eq:GR2_ent}) is monogamous for all $n$-mode Gaussian states $\rho_{A_1 A_2 \ldots A_n}$,
\begin{equation}\label{eq:mono}
\begin{array}{c}{\cal E}_2(\rho_{A_1:A_2 \ldots A_n}) - \sum_{j=2}^{n} {\cal E}_2(\rho_{A_1:A_j}) \geq 0\,,\end{array}
\end{equation}
where each $A_j$ comprises one mode only.
\end{theorem}
\noindent {\it Proof.} The structure of the proof follows closely the one for the tangle of $n$-qubit systems \cite{osborne}. It suffices to prove the inequality for tripartite Gaussian states $\rho_{A_1 A_2 A'_3}$ where $A'_3$ comprises $n-2$ modes (with $n$ arbitrary), as iterative applications to  ${\cal E}_2(\rho_{A_1:A'_3})$ would then imply Eq.~(\ref{eq:mono}). It is further enough to prove the inequality on pure states, as it would then extend to mixed ones by convexity \cite{ckw,osborne,contangle,hiroshima}. Exploiting the phase-space Schmidt decomposition \cite{ourreview,holevowerner,morejens}, when $\rho_{A_1 A_2 A'_3}$ is pure, the state of subsystem $A'_3$ is locally equivalent to a $2$-mode state, tensored by $n-4$ irrelevant vacuum modes. The problem reduces to proving that ${\cal E}_2(\rho_{A:BC}) \geq {\cal E}_2(\rho_{A:B}) + {\cal E}_2(\rho_{A:C})$ for an arbitrary pure Gaussian state $\rho_{ABC}$, where $A$ and $B$ are single modes, while $C$ groups two modes. Noting that ${\cal E}_2(\rho_{A:BC}) ={\cal S}_2(\rho_A)$, and exploiting Eq.~(\ref{eq:kw}), we see that  Eq.~(\ref{eq:mono}) is verified if one establishes that
\begin{equation}\label{eq:JE}
{\cal J}_2(\rho_{A|B}) \geq {\cal E}_2(\rho_{A:B})
\end{equation}
holds for all (mixed) two-mode Gaussian states $\rho_{AB}$. The inequality (\ref{eq:JE}) is proven in the Appendix \cite{epaps}, which concludes the proof of the Theorem. \hfill $\blacksquare$

Let us analyze in detail the case of $\rho_{A_1A_2A_3}$ being a pure $3$-mode Gaussian state, whose CM is characterized up to local unitaries by three parameters (local symplectic invariants)  $a_j \geq 1$, with ${\cal S}_2(\rho_{A_j})=\ln a_j$ ($j=1,2,3$) \cite{3modi,epaps}. We define the residual entanglement emerging from the monogamy inequality as ${\cal E}_2(\rho_{A_1:A_2:A_3}) = {\cal E}_2(\rho_{A_1:A_2A_3}) - {\cal E}_2(\rho_{A_1:A_2}) - {\cal E}_2(\rho_{A_1:A_3})$. This quantity, which can be calculated exactly \cite{contangle,3modi}, depends in general on the focus mode (say $A_1$) chosen for the decomposition of the bipartite entanglements. Remarkably, we find that for all pure three-mode Gaussian states which are fully inseparable and display entanglement in all global and reduced bipartitions (${\cal E}_2(\rho_{A_i:A_j}) > 0,\,{\cal E}_2(\rho_{A_i:A_j A_k}) > 0,\,\forall i\neq j\neq k$), the residual GR$2$ entanglement ${\cal E}_2(\rho_{A_1:A_2:A_3})$ is invariant under mode permutations, thus representing (to the best of our knowledge) the first and only known intrinsically {\it bona fide} measure of genuine tripartite Gaussian entanglement \cite{notecontangle}. We report its explicit formula here, while a derivation is provided in  \cite{epaps},
\begin{eqnarray}
\label{eq:E23}
&&{\cal E}_2(\rho_{A_1:A_2:A_3})\,\, =\,\, \ln\, (8 a_1 a_2 a_3) \\
&&\begin{array}{c}-\ln\big[-1-\sqrt{\delta}+\sum_{i=1}^3(2a_i^2-a_i^4) + \sum_{i\neq j=1}^3 (a_i^2 a_j^2)\big]\end{array}\,,
 \nonumber
\end{eqnarray}
with $\delta=\prod_{\mu,\nu=0}^{1} \big((a_1+(-1)^\mu a_2 + (-1)^\nu a_3)^2-1\big)$.
The formula (\ref{eq:E23}) holds when $|a_i-a_j|+1 <a_k < (a_i^2+a_j^2-1)^{\frac12}$.

We finally remark that, importantly, the GR$2$ discord Eq.~(\ref{eq:D2}) also turns out to be  a monogamous measure of quantum correlations for arbitrary pure three-mode Gaussian states $\rho_{A_1A_2A_3}$ (unlike the von Neumann entropy-based  discord \cite{giordazzo,adessodatta,georgie}). By using Eq.~(\ref{eq:kw}), one finds (see also \cite{koashi,fanchini}) ${\cal D}_2(\rho_{A_1:A_2:A_3})  \doteq  {\cal D}_2(\rho_{A_2|A_2A_3}) - {\cal D}_2(\rho_{A_1|A_2})-{\cal D}_2(\rho_{A_1|A_3}) = {\cal E}_2(\rho_{A_1:A_2:A_3})$. In other words, the residual tripartite discord equates the residual tripartite entanglement for pure $\rho_{A_1A_2A_3}$. This extends to the multipartite case the equivalence between entanglement and general quantum correlations valid for pure bipartite states \cite{zurek,QE,georgiemulti}, and places ${\cal D}_2$ as the only known measure of quantumness beyond entanglement in continuous variable systems that fulfills monogamy \cite{notesimilar}.


\noindent {\bf Conclusions.}---
In this Letter we planted the seeds for a full Gaussian quantum information theory \cite{GQI} using the R\'{e}nyi-$2$ entropy ${\cal S}_2$. This is possible thanks to the fact, proven in Theorem~\ref{teo}, that such an entropy satisfies the strong subadditivity inequality for arbitrary Gaussian states of quantum harmonic systems. We employed ${\cal S}_2$ to define valid measures of entanglement, total, classical, and quantum correlations, highlighting their properties. The R\'{e}nyi-$2$ mutual information is intimately related to Wigner distribution sampling by homodyne detections in phase space. The residual R\'{e}nyi-$2$ entanglement measure allows for a quantification of genuine tripartite entanglement in three-mode fully inseparable pure Gaussian states, which is invariant under mode permutations. We argue that the measures defined in this Letter should be adopted as privileged tools to address the quantification of relevant correlations in Gaussian states. A very recent application to relativistic quantum information has been reported \cite{rqi}.
%

By conception, this work has been  biased towards Gaussian states and operations.
However, the approach pursued here can be extended to arbitrary, even non-Gaussian $n$-mode states $\rho$ of continuous variable systems, provided one chooses the Wehrl  entropy \cite{wehrl} to quantify  the informational content of $\rho$. Such an entropy is operationally associated to phase-space sampling via heterodyne detections \cite{vecchiocavaliere} as it corresponds to the continuous Boltzmann--Gibbs--Shannon entropy of the Husimi $Q$ distribution of $\rho$, $Q(\gr\alpha) = \pi^{-n} \bra{\gr\alpha} \rho \ket{\gr\alpha}$, which is a valid (nonnegative) probability distribution for {\it all} quantum states \cite{zykbook}. One can then adopt a distance measure between any two $\rho_1$ and $\rho_2$ in terms of the relative Shannon entropy between their respective $Q$ distributions, define ensuing correlation measures, and so on. Such a formalism could also accommodate a measure of non-Gaussianity of quantum states \cite{genoni}, and one might then naturally compare Gaussian with non-Gaussian operations for the realization  of specific tasks \cite{giovannetti} such as extracting classical correlations \cite{gamid,wernercv}, maximizing information-disturbance trade-off \cite{mistatradeoff} and performing optimal cloning of coherent states \cite{coherentcloning}. This will be the subject of further study.

\noindent {\bf Acknowledgments.}--- {We acknowledge enlightening discussions with I. Apicella, V. P. Belavkin, N. Cerf, F. P. Cimino Wood, A. Datta, N. Datta, J. Eisert, V. Giovannetti, M. Guta, M. van Horssen,  L. Mi\v{s}ta Jr., M. Mosonyi,  S. Pirandola, A. Winter. GA thanks the University of Nottingham for financial support through an Early Career Research and Knowledge Transfer Award. AS thanks  Jessica \& Dave for moral and financial support.}


\clearpage
\begin{widetext}
\begin{center}
Supplemental Material

\medskip

{\large{\bf Measuring Gaussian Quantum Information and Correlations Using the R\'{e}nyi Entropy of Order $\boldsymbol 2$}}

\medskip

Gerardo Adesso, Davide Girolami, and Alessio Serafini
\end{center}
\appendix
\setcounter{equation}{0}
\setcounter{page}{1}

\section{Derivation of formulae (\ref{eq:sampling}) and (\ref{eq:relent})}

For the sake of completeness, here we provide the reader with an explicit derivation of Eqs.~(\ref{eq:sampling}) and (\ref{eq:relent}),
on which the operational interpretation of the R\'{e}nyi-$2$ entropy is ultimately based.

Direct substitution of Eq.~(\ref{eq:wigner}) into (\ref{eq:sampling}) yields
\be
H(W_{\rho})=\int_{{\mathbbm R}^{2n}}\frac{1}{\pi^n \sqrt{\det{\gam}}} \exp\big(-\boldsymbol{\xi}^{\sf T} \boldsymbol{\gamma}^{-1} \boldsymbol{\xi}\big)
\left[\boldsymbol{\xi}^{\sf T} \boldsymbol{\gamma}^{-1} \boldsymbol{\xi} + n\ln\pi+\frac12\ln\left({\rm det}{\gr \gamma}\right)\right]
{\rm d}^{2n}{\gr \xi} \; . \label{eq:expli}
\ee
The first term in the square bracket is conveniently handled by performing the integration in phase space coordinates ${\gr \xi}$
that diagonalize the symmetric, positive definite matrix ${\gr \gamma}$ (whose eigenvalues will be denoted by $\gamma_{j}$), and by noting that, for any $\gamma>0$, one has
\be
\frac1{\sqrt{\pi\gamma}}\int_{-\infty}^{+\infty} \exp(-\gamma^{-1} x^2) x^2 {\rm d}x = \frac{\gamma}{2} \; . \label{eq:gauint}
\ee
Hence, by applying (\ref{eq:gauint}) as well as the normalization of the Wigner function to (\ref{eq:expli}), one obtains Eq.~(\ref{eq:sampling}):
\be
H(W_{\rho}) = \sum_{j=1}^{2n} \frac{\gamma_j}{2\gamma_j} +n\ln\pi + S_2(\rho) = n+n\ln\pi + S_2(\rho) \; .
\ee

Next, the substitution of (\ref{eq:wigner}) into (\ref{eq:relent}) leads to:
\begin{eqnarray}
H(W_{\rho_1}\|W_{\rho_2}) = -H(W_{\rho_1})
+ \int_{{\mathbbm R}^{2n}}\frac{1}{\pi^n \sqrt{\det{\gam_1}}} \exp\big(-\boldsymbol{\xi}^{\sf T} \boldsymbol{\gamma}_1^{-1} \boldsymbol{\xi}\big)
\left[\boldsymbol{\xi}^{\sf T} \boldsymbol{\gamma}_2^{-1} \boldsymbol{\xi} + n\ln\pi+\frac12\ln\left({\rm det}{\gr \gamma}_2\right)\right]
{\rm d}^{2n}{\gr \xi} \, .
\end{eqnarray}
Once again, it is expedient to carry out the integration in phase space coordinates where ${\gr \gamma}_2$ is diagonal
(with eigenvalues $\gamma_{2,j}$, while the entries of $\gam_1$ will be denoted by $\gamma_{1,jk}$). Then, Eq.~(\ref{eq:gauint})
and the normalization of $W_{\rho_1}$ lead to (\ref{eq:relent})
\be
H(W_{\rho_1}\|W_{\rho_2}) = \frac12 \ln\left(\frac{\det \gam_2}{\det \gam_1}\right) - n + \sum_{j=1}^{2n}
\frac{\gamma_{1,jj}}{2\gamma_{2,j}} = \frac12 \ln\left(\frac{\det \gam_2}{\det \gam_1}\right) - n + \frac12{\rm tr}(\gam_1\gam_2^{-1}) \; ,
\ee
where the last step follows from the invariance of the quantity ${\rm tr}(\gam_1\gam_2^{-1})$ under changes of basis.


\section{Explicit formulae for two-mode Gaussian correlations}

\bigskip

\noindent{\bf Standard form.}---
The CM $\gam_{AB}$ of any two-mode Gaussian state $\rho_{AB}$ can be transformed, by means of local unitary operations, into a standard form of the type \cite{ourreview}
\begin{equation}
\label{eq:gamsf}
\gam_{AB}=\left(\begin{array}{cc}
{\gam_A}&{\gr\varsigma_{AB}}\\
{\gr\varsigma}_{AB}^{\sf T}&{\gam_B}
\end{array}\right) = \left(\begin{array}{cccc}
a&0&c_{+}&0\\
0&a&0&c_{-}\\
c_{+}&0&b&0\\
0&c_{-}&0&b
\end{array}\right)\;,
\end{equation}
where $a,b\geq 1$, $\left[\left(a^2-1\right) \left(b^2-1\right)-2 c_- c_+-a b c_+^2+c_-^2 \left(-a b+c_+^2\right)\right] \geq 0$, and we can set $c_+ \ge |c_-|$ without losing any generality. These conditions ensure that the uncertainty relation  $\gam_{AB} \geq i \gr\omega^{\oplus 2}$ is verified, which is a {\it bona fide} requirement for the CM $\gam_{AB}$ to be associated with a physical Gaussian state in a two-mode infinite-dimensional Hilbert space \cite{ourreview}.
Recall that for pure Gaussian states, $b=a$, $c_+=-c_-=\sqrt{a^2-1}$.

All the formulae presented in the following will be written explicitly for standard form CMs for simplicity. However, they can be recast in a locally invariant form by expressing them in terms of the four local symplectic invariants of a generic two-mode Gaussian state \cite{serafozzididattico}, $I_1=\det \gam_A$, $I_2=\det \gam_B$, $I_3=\det{\gr\varsigma}_{AB}$, $I_4=\det\gam_{AB}$. This is accomplished by inverting the relations $I_1=a^2, I_2=b^2, I_3=c_+ c_-, I_4=(a b -c_+)(a b - c_-)$ so that the $\{I_j\}_{j=1}^4$ appear explicitly in the formulae below \cite{ourreview}. The obtained expressions would then be valid for two-mode CMs in any symplectic basis, beyond the standard form.

\bigskip

\noindent {\bf GR$\gr 2$ entanglement.}---
For generally mixed two-mode Gaussian states $\rho_{AB}$, the R\'{e}nyi-$2$ entanglement measure ${\cal E}_2(\rho_{A:B})$, defined by Eq.~(\ref{eq:GR2_ent}) in the main text, admits the following expression if the CM $\gam_{AB}$ is in standard form
\cite{geof,ordering},
\begin{equation}\label{eq:E2AB}
{\cal E}_2(\rho_{A:B}) = \frac12 \ln \left(\inf_{\theta \in [0,2\pi]}  m_\theta (a,b,c_+, c_-)\right)\,,
\end{equation}
with
\begin{eqnarray}\label{eq:mfunc}
m_\theta (a,b,c_+, c_-)\ =\ 1 &+&
\left[c_+(ab-c_-^2)-c_-+\cos \theta \sqrt{\left[a -
b(ab-c_-^2)\right]\left[b-a(ab-c_-^2)\right]}\right]^2
\nonumber \\
& \times & \left\{
2\left(ab-c_-^2\right)\left(a^2+b^2+2c_+c_- \right) +\ \sin \theta\left(a^2-
b^2\right)\sqrt{1-\frac{\left[c_+(ab-c_-^2)+c_-\right]^2}{\left[a -
b(ab-c_-^2)\right]\left[b-a(ab-c_-^2)\right]}}\right. \nonumber \\
& &\left.\ -\ \frac{\cos \theta\left[2abc_-^3+\left(a^2+
b^2\right)c_+c_-^2+\left(\left(1-2b^2\right)a^2+
b^2\right)c_--ab\left(a^2+b^2- 2\right)c_+\right]}{\sqrt{\left[a -
b(ab-c_-^2)\right]\left[b-a(ab-c_-^2)\right]}} \right\}^{-1}\,.
\end{eqnarray}
The optimal $\theta$ minimizing Eq.~(\ref{eq:mfunc}) can be found numerically for general two-mode Gaussian states \cite{geof}, and analytically for relevant subclasses of states (including symmetric states \cite{eofg}, squeezed thermal states, and so-called GLEMS---Gaussian states of partial minimum uncertainty \cite{ordering}).

\bigskip

\noindent {\bf GR$\gr 2$ classical correlations and discord.}---
For generally mixed two-mode Gaussian states $\rho_{AB}$, the R\'{e}nyi-$2$ measures of one-way classical correlations ${\cal J}_2(\rho_{A|B})$ and quantum discord ${\cal D}_2(\rho_{A|B})$, defined by Eqs.~(\ref{eq:J2}) and (\ref{eq:D2}) in the main text, respectively, admit the following expression if the CM $\gam_{AB}$ is in standard form \cite{adessodatta}
\begin{eqnarray}
{\cal J}_2(\rho_{A|B}) &=& \ln a -    \frac12 \ln \left(\inf_{\lambda,\varphi}{\det\tilde{\gr\gamma}^{\Pi_{\lambda,\varphi}}_A}\right)\,, \label{eq:J2AB} \\
{\cal D}_2(\rho_{A|B}) &=& \ln b - \frac12 \ln \big(\det{\gam_{AB}}\big) +   \frac12 \ln \left(\inf_{\lambda,\varphi}{\det\tilde{\gr\gamma}^{\Pi_{\lambda,\varphi}}_A}\right)\,, \label{eq:D2AB}
\end{eqnarray}
with $\lambda \in (0,\infty),\,\varphi\in[0,2\pi]$, and
\begin{equation}\label{eq:detcond}
\det\tilde{\gr\gamma}^{\Pi_{\lambda,\varphi}}_A =  \frac{2 a^2 (b + \lambda) (1 + b \lambda)-a \left(c_+^2+c_-^2\right) \left(2 b \lambda +\lambda ^2+1\right) +2 c_+^2 c_-^2 \lambda +a \left(c_+^2-c_-^2\right) \left(\lambda ^2-1\right) \cos (2 \varphi)}{{2 (b + \lambda) (1 + b \lambda)}}\,.
\end{equation}
The optimal values of $\lambda$ and $\varphi$ minimizing Eq.~(\ref{eq:detcond}) can be found analytically\footnote{
In general, given that ${\cal S}_2(\Lambda^\Pi_B)$ is concave on the convex hull of the set of Gaussian states ---having inherited the property from the Shannon entropy of the corresponding Wigner distributions, via Eq.~(\ref{eq:sampling})--- and given that every Gaussian state admits  a convex decomposition into pure Gaussian states (see \cite{serafozzinazi,adessodatta,gamid} for more details), it follows that the optimizations in Eqs.~(\ref{eq:J2},\ref{eq:D2}) for an arbitrary number of modes are always achieved by pure Gaussian seed elements, i.e., $\det(\gr\Gamma_B^\Pi)=1$. This simplifies considerably the evaluation of GR$2$ one-way classical correlations and discord.} for all two-mode Gaussian states \cite{adessodatta}. In particular, for standard form CMs, one gets
\begin{equation}\label{eq:optdetcond}
\inf_{\lambda,\varphi}\det\tilde{\gr\gamma}^{\Pi_{\lambda,\varphi}}_A =
\left\{
\begin{array}{c}
 a \left(a-\frac{c_+^2}{b}\right)\,,\qquad \qquad \qquad \text{if}\quad \left(a b^2 c_-^2-c_+^2 \left(a+b c_-^2\right)\right) \left(a b^2 c_+^2-c_-^2 \left(a+b c_+^2\right)\right)<0\,; \\
 \frac{2 \left|c_- c_+\right| \sqrt{\left(a \left(b^2-1\right)-b c_-^2\right) \left(a \left(b^2-1\right)-b c_+^2\right)}+\left(a \left(b^2-1\right)-b c_-^2\right) \left(a \left(b^2-1\right)-b c_+^2\right)+c_-^2 c_+^2}{\left(b^2-1\right)^2} \,,\qquad \text{otherwise.}
\end{array}
\right.
\end{equation}
Inserting Eq.~(\ref{eq:optdetcond}) into Eqs.~(\ref{eq:J2AB},\ref{eq:D2AB}) one gets closed formulae for the one-way GR$2$ classical correlations and for the GR$2$ discord of general two-mode Gaussian states.

\bigskip

\section{Proof that GR$\gr 2$ classical correlations exceed GR$\gr 2$ entanglement}
Here we prove that the inequality
\begin{equation}\label{eq:JEpaps}
{\cal J}_2(\rho_{A|B}) \geq {\cal E}_2(\rho_{A:B})\,,
\end{equation}
holds for all two-mode Gaussian states $\rho_{AB}$. This is a central step in the proof of the general monogamy inequality for the GR$2$ entanglement measure, reported in Theorem~\ref{teomono} in the main text (see also \cite{osborne}).
Without loss of generality, we can assume the CM $\gam_{AB}$ in standard form.

We observe from Eqs.~(\ref{eq:E2AB},\ref{eq:J2AB}) that setting any value of $\theta$ in (\ref{eq:mfunc}) provides an upper bound to ${\cal E}_2$, while setting any value of $\lambda,\varphi$ in (\ref{eq:detcond}) provides a lower bound to ${\cal J}_2$,
\begin{eqnarray*}
{\cal E}_2(\rho_{A:B}) &\le&  \frac12 \ln \big(m_\theta (a,b,c_+, c_-)\big)\,, \\
{\cal J}_2(\rho_{A|B}) &\ge&  \ln a -    \frac12 \ln \left({\det\tilde{\gr\gamma}^{\Pi_{\lambda,\varphi}}_A}\right)\,,
\end{eqnarray*}
We will then set $\theta=\pi$ (similarly to what done in \cite{hiroshima}), and $\lambda=1,\varphi=0$ (corresponding to heterodyne detections on $B$ \cite{giordazzo}), and proceed to prove that $\ln a -    \frac12 \ln \left({\det\tilde{\gr\gamma}^{\Pi_{1,0}}_A}\right) \geq
\frac12 \ln \big(m_\pi (a,b,c_+, c_-)\big)$, which implies (\ref{eq:JEpaps}).
We then want to prove the inequality $1+F^2/G \leq J$, where
\begin{eqnarray*}
F&=&c_+(ab-c_-^2)-c_-+ \sqrt{\left[a -
b(ab-c_-^2)\right]\left[b-a(ab-c_-^2)\right]}\,, \\
G&=&2\left(ab-c_-^2\right)\left(a^2+b^2+2c_+c_+ \right)  -\ \frac{2abc_-^3+\left(a^2+
b^2\right)c_+c_-^2+\left(\left(1-2b^2\right)a^2+
b^2\right)c_--ab\left(a^2+b^2- 2\right)c_+}{\sqrt{\left[a -
b(ab-c_-^2)\right]\left[b-a(ab-c_-^2)\right]}}\,,\\
J&=&\frac{a^2(1+b)^2}{(a+a b-c_+^2)(a+a b-c_-^2)}\,.
\end{eqnarray*}
This is equivalent to show that, defining $K=K(a,b,c_+,c_-)\doteq J-F^2/G-1$, we have $\min_{(a,b,c_+,c_-)} K= 0$.

Let us search for the absolute minimum of $K$. This function has no singularities apart from the trivial case of two-mode vacuum states ($a=b=1,c_\pm=0$) for which we know that $\lim_{(a,b,c_+,c_-)\rightarrow{(1,1,0,0)}}K=0$. Thus, we can focus on the stationary points of $K$ and on the values at the boundaries of its domain. We adopt the Karush-Kuhn-Tucker method \cite{karush,tucker}. Given constraints $f_i=f_i(a,b,c_+,c_-) \leq 0 $ and associated multipliers $\Lambda_i \geq 0$, the coordinates  $(a^*,b^*,c_+^*,c_-^*,\Lambda_i^*)$ associated to a minimum of $K$ satisfy the following conditions:
\begin{itemize}
\item[(a)]$\left(\gr\nabla, \frac{\partial}{\partial \Lambda_i} \right)L(a,b,c_+,c_-, \Lambda_i)|_{(a^*,b^*,c_+^*,c_-^*,\Lambda_i^*)}=\left(\gr\nabla,\frac{\partial}{\partial \Lambda_i}\right)(K+\sum_i \Lambda_i f_i)|_{(a^*,b^*,c_+^*,c_-^*,\Lambda_i^*)}=0  $
\item[(b)]$\Lambda_i f_i(a^*,b^*,c_+^*,c_-^*)=0, \quad  \Lambda_i \geq 0\  \ \forall i$
\item[(c)] there exists  a vector $ \vec{x}=\{x_i\}$ such that $ (\gr\nabla K-\sum_i x_i \gr\nabla f_i)|_{(a^*,b^*,c_+^*,c_-^*)}=0.$
\end{itemize}
Conditions (a) and (b) are necessary, while condition (c) is sufficient.
 In this case, we have
 \begin{eqnarray*}
 f_1&=&a^2+b^2 +2 c_+ c_--1-(a b-c_+^2)(a b-c_-^2)\,, \\
 f_2&=&-a^2-b^2+ 2 c_+ c_-+1+(a b-c_+^2)(a b-c_-^2)\,, \\
 f_3&=&-c_+-c_-\,.
 \end{eqnarray*}
 A bit of algebra reveals that condition (a) is verified by the solutions to the following system
 \begin{eqnarray}
 \left\{ \begin{array}{l}
 \displaystyle\frac{\partial K}{\partial a} \frac{\partial f_1}{\partial b}  =\frac{\partial K}{\partial b}\frac{\partial f_1}{\partial a}   \\ \\
\displaystyle\frac{\partial K}{\partial c_-} = \frac{\partial K}{\partial c_+}+\frac{\frac{\partial K}{\partial a}(\frac{\partial f_1}{\partial c_-}-\frac{\partial f_1}{\partial c_+})}{\frac{\partial f_1}{\partial a}}+\Lambda_2 \left(\frac{\partial f_1}{\partial c_+}+\frac{\partial f_2}{\partial c_+}-\frac{\partial f_1}{\partial c_-}-\frac{\partial f_2}{\partial c_-}\right). \\
 \end{array}
   \right.         \end{eqnarray}
 It follows that,  among the stationary points, the absolute minimum of $K$ is again reached at $a=b=1, c_\pm=0$, yielding $K=0$: it is immediate to verify that the conditions (b) and (c) are satisfied as well at $(1,1,0,0)$.  Finally, we have to check the values of $K$ at the boundaries of its domain.   The relevant cases are $a\rightarrow \infty , b\rightarrow \infty, c_+=\sqrt{a b-1},c_+=|c_-|,c_-=0$, and somewhat tedious yet straightforward analysis reveals that $K\geq 0$ always holds for all those cases.

  We therefore conclude that the function $K$ is always nonnegative, which proves Eq.~(\ref{eq:JEpaps}), thereby proving that the entanglement measure ${\cal E}_2$ satisfies the general monogamy inequality [Eq.~(\ref{eq:mono}) of the main text] for all $n$-mode Gaussian states of continuous variable quantum systems. \hfill $\blacksquare$

\bigskip

\section{Residual tripartite GR$\gr 2$ entanglement for pure three-mode Gaussian states}

Up to local unitaries, the CM $\gam_{A_1A_2A_3}$ of any pure three-mode Gaussian state can be written in the following standard form \cite{3modi}
 \begin{eqnarray}
 \gam_{A_1A_2A_3}=\left( \begin{array}{cccccc}
 a_1 & 0 &c_3^+&0&c_2^+&0\\
0 & a_1 & 0&c_3^-&0&c_2^-\\
c_3^+& 0 &a_2&0&c_1^+&0\\
0 & c_3^- & 0&a_2&0&c_1^-\\
c_2^+ & 0 &c_1^+&0&a_3&0\\
0 & c_2^-& 0&c_1^-&0&a_3\\
 \end{array}
   \right)
\end{eqnarray}
where
\begin{eqnarray*}
&& c_i^\pm = \frac{ \sqrt{[(a_i-1)^2-(a_j-a_k)^2][(a_i+1)^2-(a_j-a_k)^2]}\pm\sqrt{[(a_i-1)^2-(a_j+a_k)^2][(a_i+1)^2-(a_j+a_k)^2]}}{4\sqrt{a_j a_k}}\,, \\
&& \hbox{and } |a_j - a_k|+1\leq a_i\leq a_j+a_k-1\,,
\end{eqnarray*}
with $\{i,j,k\}$ being all possible permutations of $\{1,2,3\}$.

The GR$2$ entanglement in the two-mode reduced state with CM $\gam_{A_i A_j}$ is
\begin{equation}
{\cal E}_2(\rho_{A_i:A_j}) = \frac12 \ln g_k \,,
\end{equation}
with \cite{ordering}
\begin{equation}\label{eq:mglems}
g_k = \left\{
                  \begin{array}{ll}
                    1, & \hbox{if $a_k \geq \sqrt{a_i^2+a_j^2-1}$;} \\
                    \displaystyle\frac{\beta}{8 a_k^2}, & \hbox{if $\alpha_k < a_k < \sqrt{a_i^2+a_j^2-1}$;} \\
                    \displaystyle\left(\frac{a_i^2-a_j^2}{a_k^2-1}\right)^2, & \hbox{if $a_k \leq \alpha_k$.}
                  \end{array}
                \right.
\end{equation}
Here we have set
\begin{eqnarray*}
\alpha_k&=&\sqrt{\frac{2(a_i^2+a_j^2)+(a_i^2-a_j^2)^2+|a_i^2-a_j^2|\sqrt{(a_i^2-a_j^2)^2+8(a_i^2+a_j^2)}}{2(a_i^2+a_j^2)}}\,,\\
\beta&=&-1+2 a_1^2+2 a_2^2+2 a_3^2+2 a_1^2 a_2^2+2 a_1^2 a_3^2+2 a_2^2 a_3^2-a_1^4-a_2^4-a_3^4-\sqrt{\delta}\,,\\
\delta&=&(-1+a_1-a_2-a_3) (1+a_1-a_2-a_3) (-1+a_1+a_2-a_3) (1+a_1+a_2-a_3) \\ &\times& (-1+a_1-a_2+a_3) (1+a_1-a_2+a_3) (-1+a_1+a_2+a_3) (1+a_1+a_2+a_3)\,.
\end{eqnarray*}
The residual GR$2$ entanglement, with respect to the focus mode $A_i$, is
\begin{eqnarray}
{\cal E}_2(\rho_{A_i:A_j:A_k})&=& {\cal E}_2(\rho_{A_i:A_jA_k}) - {\cal E}_2(\rho_{A_i:A_j}) - {\cal E}_2(\rho_{A_i:A_k}) \nonumber \\
&=& \frac12 \ln \left(\frac{a_i^2}{g_k\ g_j}\right)\,.
\end{eqnarray}
In general, this expression is dependent on the choice of the focus mode. However, let us consider the relevant case of a fully inseparable three-mode pure Gaussian state, for which entanglement is nonzero for all global splittings and for all reduced two-mode bipartitions, ${\cal E}_2(\rho_{A_i:A_jA_k})>0,\,{\cal E}_2(\rho_{A_i:A_j})>0,\,\forall \{i,j,k\}$. In our parametrization, this occurs when \cite{ordering} \begin{equation}
|a_i-a_j|+1<a_k < \sqrt{a_i^2+a_j^2-1}\,,\end{equation} for all mode permutations. It is immediate to see that the simultaneous verification of such a condition for all mode permutations imposes $a_k > \alpha_k\,\forall k=1,2,3$. In this case, exploiting Eq.~(\ref{eq:mglems}), the residual GR$2$ entanglement becomes
\begin{equation}
{\cal E}_2(\rho_{A_i:A_j:A_k}) = \frac12 \ln \left( \frac{64 a_i^2 a_j^2 a_k^2}{\beta^2} \right)\,,
\end{equation}
which is manifestly invariant under mode permutations, as reported in Eq.~(\ref{eq:E23}) in the main text.
This symmetry is broken on states for which some of the reduced two-mode bipartitions become separable.

\clearpage
\end{widetext}


\begin{thebibliography}{99}
\bibitem{footnotedatta}
In particular, the only classically correlated Gaussian states $\rho_{AB}$ (with ${\cal D}_2(\rho_{A|B})=0$) are those that saturate the strong subadditivity inequality (\ref{eq:ss2}) on a tripartite Gaussian state $\rho_{ABC}$ where $C$ is an ancillary system realizing the measurements on $B$, as characterized in \cite{adessodatta,petz}.
\end{thebibliography}
\end{document}